\documentclass[a4paper,12pt,number]{elsarticle}

\usepackage{amssymb,amsmath}
\usepackage{graphicx}
\usepackage{hyperref}
\usepackage{geometry}
\usepackage{xcolor}
\usepackage[normalsize]{subfigure}
\usepackage{lineno}
\usepackage{array}
\usepackage{ulem}
\usepackage{hyperref}
\usepackage{nomencl}

\newcommand{\save}[1]{}

\newcommand{\rem}[1]{}

\newcommand{\reftab} [1]{Tab.~\ref{tab:#1}}
\newcommand{\refsec} [1]{Section~\ref{sec:#1}}
\newcommand{\reffig} [1]{Fig.~\ref{fig:#1}}
\newcommand{\reffigs}[2]{Figs~\ref{fig:#1} and~\ref{fig:#2}}

\newcommand{\tableset}[2]{\renewcommand{\arraystretch}{#1} \setlength\tabcolsep{#2pt}}
\newcommand{\tm}{\leavevmode\hbox{$\rm {}^{TM}$}\,}

\journal{arxiv}

\makenomenclature

\begin{document}

\begin{frontmatter}
  
\title{A framework for integrated design of algorithmic architectural forms}

\author[1]{Ladislav Svoboda\corref{cor}}
\ead{ladislav.svoboda@fsv.cvut.cz}
\cortext[cor]{Corresponding author. Tel.:~+420-224-354-495}
\author[1,2]{Jan Nov\'{a}k}
\ead{novakj@cml.fsv.cvut.cz}
\author[3]{Luk\'{a}\v{s} Kurilla}
\ead{mail@kurilluk.com}
\author[1]{Jan Zeman}
\ead{zemanj@cml.fsv.cvut.cz}

\address[1]{Department of Mechanics, Faculty of Civil Engineering, Czech Technical University in Prague, Th\'{a}kurova 7, \mbox{166 29 Praha 6}, Czech Republic}
\address[2]{Institute of Structural Mechanics, Faculty of Civil Engineering, Brno University of Technology}
\address[3]{Department of Construction Engineering I, Faculty of Architecture, Czech Technical University in Prague, Th\'{a}kurova 9, \mbox{166 34 Praha 6}, Czech Republic}

\begin{abstract}
%
 This paper presents a methodology and software tools for parametric design of complex architectural objects, called digital or algorithmic forms. In order to provide a flexible tool, the proposed design philosophy involves two open source utilities Donkey and MIDAS written in Grasshopper algorithm editor and C++, respectively, that are to be linked with a scripting-based architectural modellers Rhinoceros, IntelliCAD and the open source Finite Element solver OOFEM. The emphasis is put on the mechanical response in order to provide architects with a consistent learning framework and an insight into structural behaviour of designed objects. As demonstrated on three case studies, the proposed modular solution is capable of handling objects of considerable structural complexity, thereby accelerating the process of finding procedural design parameters from orders of weeks to days or hours.
\end{abstract}


\end{frontmatter}


\section{Introduction}
%
Architecture, the essential component to urban environment, impacts every individual, which, in turn, influences back the next generation of art. In each period of civilisation, architecture has reflected the level of societal progress by integrating the state of the art from various fields of human activity. In other words, we can understand architecture as a multidisciplinary subject combining current knowledge not only from technical fields but also from Humanities, Ecology or Military and defence. However, the increasing level of knowledge characterised by narrow specialisation results in educational institutions producing architects unprepared for a strong cross-disciplinary dialogue vital in today's complex society~\cite{MULTIARCH}.

\begin{figure}[t]
  \centering
  \begin{tabular}{c@{\hspace{7mm}}c}
    \includegraphics*[width=45 mm]{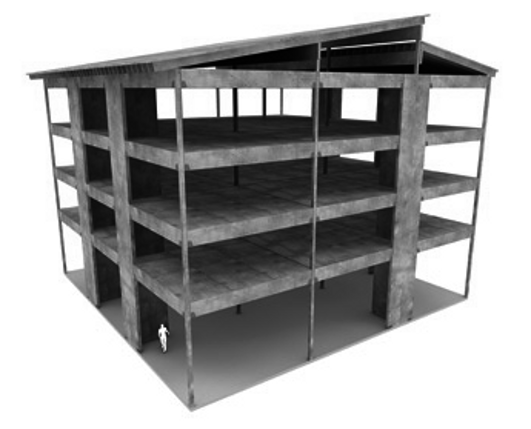} &
    \includegraphics*[width=63 mm]{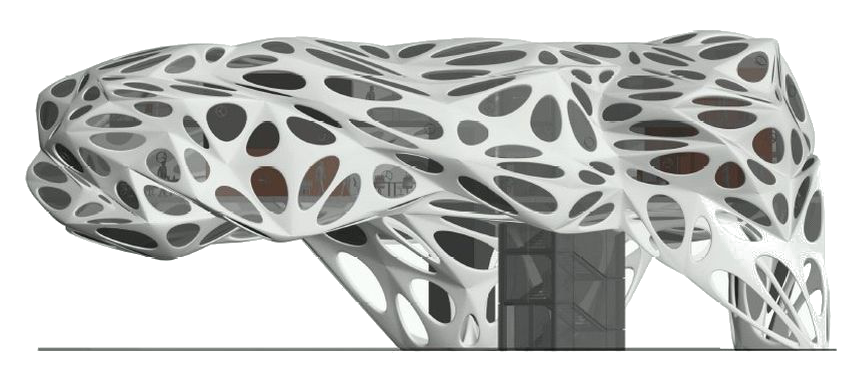} \\
    a) & b)
  \end{tabular}
  \caption{Comparison of a) architrave and b) amorphous forms.}
  \label{fig:arch-newold}
\end{figure}

The lack of discussion and mutual understanding is evident especially between architects and structural engineers as mentioned in the work of Clune et al.~\cite{clune2012object}. As long as the architects traditional, especially architrave, structures where the dimensions of particular components were the only unknowns, see~\reffig{arch-newold}a, a complete structural assessment could be performed at late stages of the design process. Current designers, however, often employ sophisticated computer aided environments to generate complex amorphous light-weight forms, thereby requiring a conceptual static assessment already at the beginning of the design process, see~\reffig{arch-newold}b.

\subsection{BIM concept}
%
To start our discussion on interdisciplinary cooperation, let us first recall the Building Information Modelling (BIM) concept~\cite{BIM}. BIM is a recent and popular way of managing complex collaboration and communication processes among architects, structural engineers and construction industry members. The term BIM involves the process of generating and managing building data throughout the life cycle of a structure. The result is a data-rich, object-based, usually three-dimensional ``Building Information Model'' created by specialised CAD-BIM systems. It integrates all the information on the construction from architectural design (geometry of building elements, spatial relations as connectivity, etc.), structural design (project design documentation, structural scheme) to the process of construction and maintenance (detailed design, building process and/or rehabilitation). Thanks to this, architects and structural engineers (and also builders and owners) can effectively generate and coordinate complex digital documentation of the structure at any phase of its existence.

\begin{figure}[t]
  \centering
  \begin{tabular}{c@{\hspace{4mm}}c}
    \includegraphics*[width=73 mm]{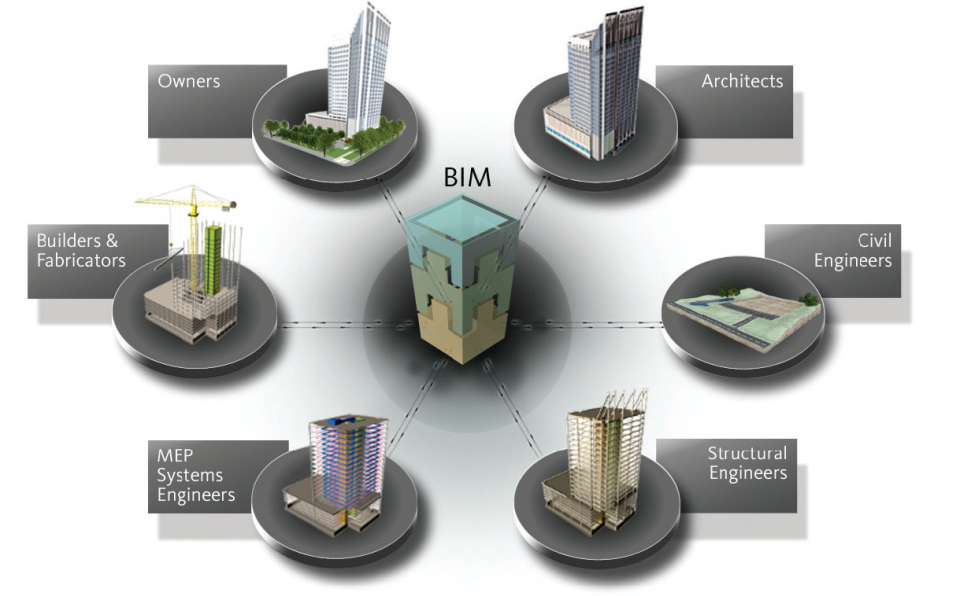} &
    \includegraphics*[width=70 mm]{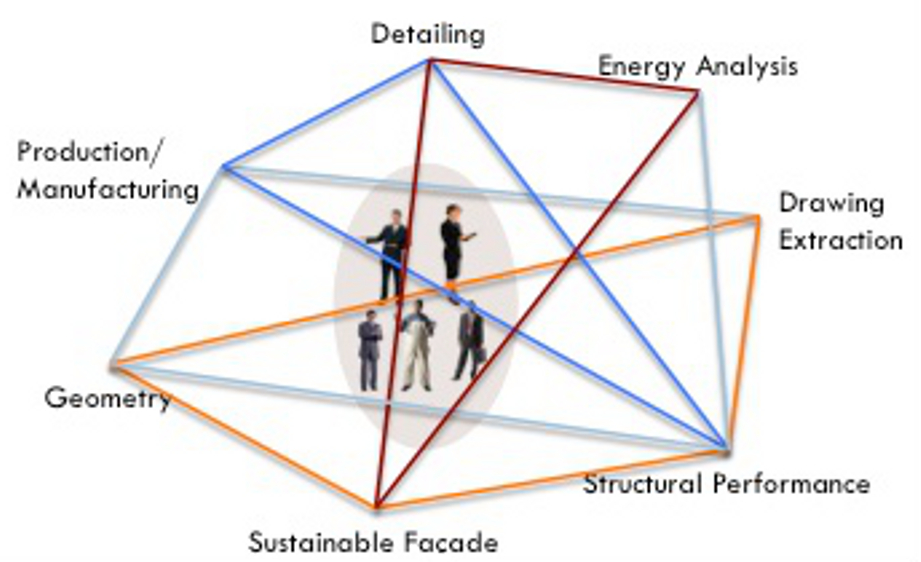} \\
    a) & b)
  \end{tabular}
  \caption{Illustration of a) BIM and b) multidisciplinary approach (courtesy of Tuba Kocaturk: BIM conference in Prague).}
  \label{fig:approach}
\end{figure}

Despite the obvious advantages, BIM only connects participants of the building industry by means of a database-like communication channel,~\reffig{approach}a. Each participant, however, remains highly specialised in his own field. This is inconsistent with our aim to enhance multidisciplinary approach in the design process, where the integration of professions into architecture should follow from exchange of mutual knowledge as sketched in~\reffig{approach}b. Moreover the BIM approach is in certain cases too cumbersome. This happens namely in initial stages of the project (investor's plan, architectural study), which is the most creative phase of the design process, taking place in a close cooperation between an architect and his client. In this case, BIM is unnecessarily complicated and general. On the other hand, this phase can last for several months (even years in extreme) and involve considerable costs. It is therefore desirable to validate the starting form at minimal time, while avoiding severe violations of structural principles.

\section{Methodology}
%
The methodology and software tools presented in this paper aim to improve the interaction between designers and structural engineers in the critical phase of the conceptual design. At this time, architects are sorting out preliminary visions penetrated by investor's plans. Resulting functional and spatial contexts may be difficult to understand even for the members of the comunity, structural engineers let alone. In these regards, we identify three goals summarized next.\\
(i) {\bf collaboration}: The developed interface can be understood as a generic tool which combines geometric modellers and a software for structural analysis~\cite{SHRNUTI}. A significant emphasis is given to the modular approach that enables the connection among arbitrary open source and commercial software packages. This strategy significantly broadens the applicability of each single module, namely, in comparison with recently developed products based on a monolithic solution, e.g.~\cite{FESCD, SandS, karamba}. In addition, the set of our tools is released under public license regulations and is freely available\footnote{www.igend.cz} to corporate and non-profit bodies. \\
%
(ii) {\bf learn}: From the viewpoint of a designer, the tools are integrated into his favourite modeller as a plug-in. It allows for structural analyses of different difficulty up to the complexity of a target artwork. Probably most importantly, the basic interface (GUI) has to be easy to use in order to not discourage a user at the first impression. As a result, the software allows the user to understand what he does rather then to provide him with plain answers on static admissibility of the structure. \\
(iii) {\bf form-finding}:
In the case of computationally less demanding structures, the analysis runs interactively. The response of the model to loading or geometry changes is visualised in the real time. This, in  combination with procedural modelling, enables relatively fast generation of a large number of variants and instant structural assessment for intuitive shaping of the structure. If necessary, such process can be automated by Evolutionary Structural Optimisation (ESO) optimisation methods, see~\cite{holzer2005design,burry2005dynamical}.










\subsection{Object-oriented model}\label{sec:object}
%
As indicated above, there is a fundamental incompatibility in cooperation between the architects and structural engineers in terms of priorities imposed on the computer model of designed objects. While architects emphasise the aesthetics aspects, structural engineers give the focus on the load-carrying system. An analysis directly performed on architectural models seams to be the most direct way. However, the complex three-dimensional CAD data are often computationally prohibitive. Moreover, a comprehensive analysis on somehow provisional inputs may easily come out uneconomic. It is thus desirable to simplify these models, while maintaining their essential structural characteristics. Typically, such a transformation is performed by a structural engineer on the basis of his experience and professional knowledge, since the full automation of the process is very difficult even with the conversion techniques developed within BIM technology.

As mentioned above our focus is on modelling of preliminary layouts. Thus, architects should tolerate simplified models, representing only the load-bearing components of the structure. If so,
the conversion into the computational model can be carried out directly in the geometric modeller with only a minimum expert intervention. In the parametric modellers, this is best achieved by exploiting their inherent scripting capabilities.

%
\begin{figure}[h!]
  \centering
  \begin{tabular}{c@{\hspace{8mm}}c@{\hspace{8mm}}c}
    \includegraphics*[height=40 mm]{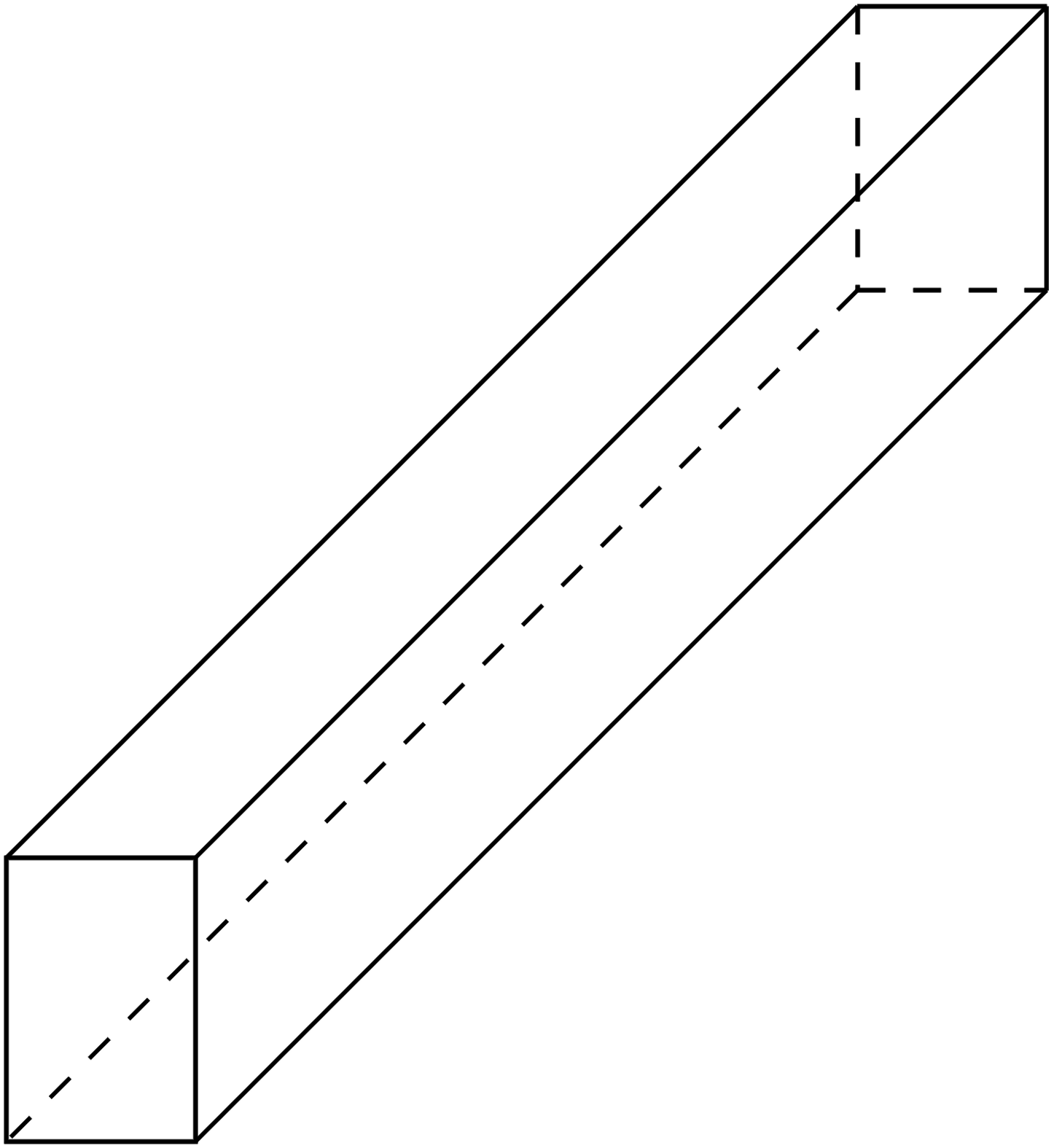} &
    \includegraphics*[height=40 mm]{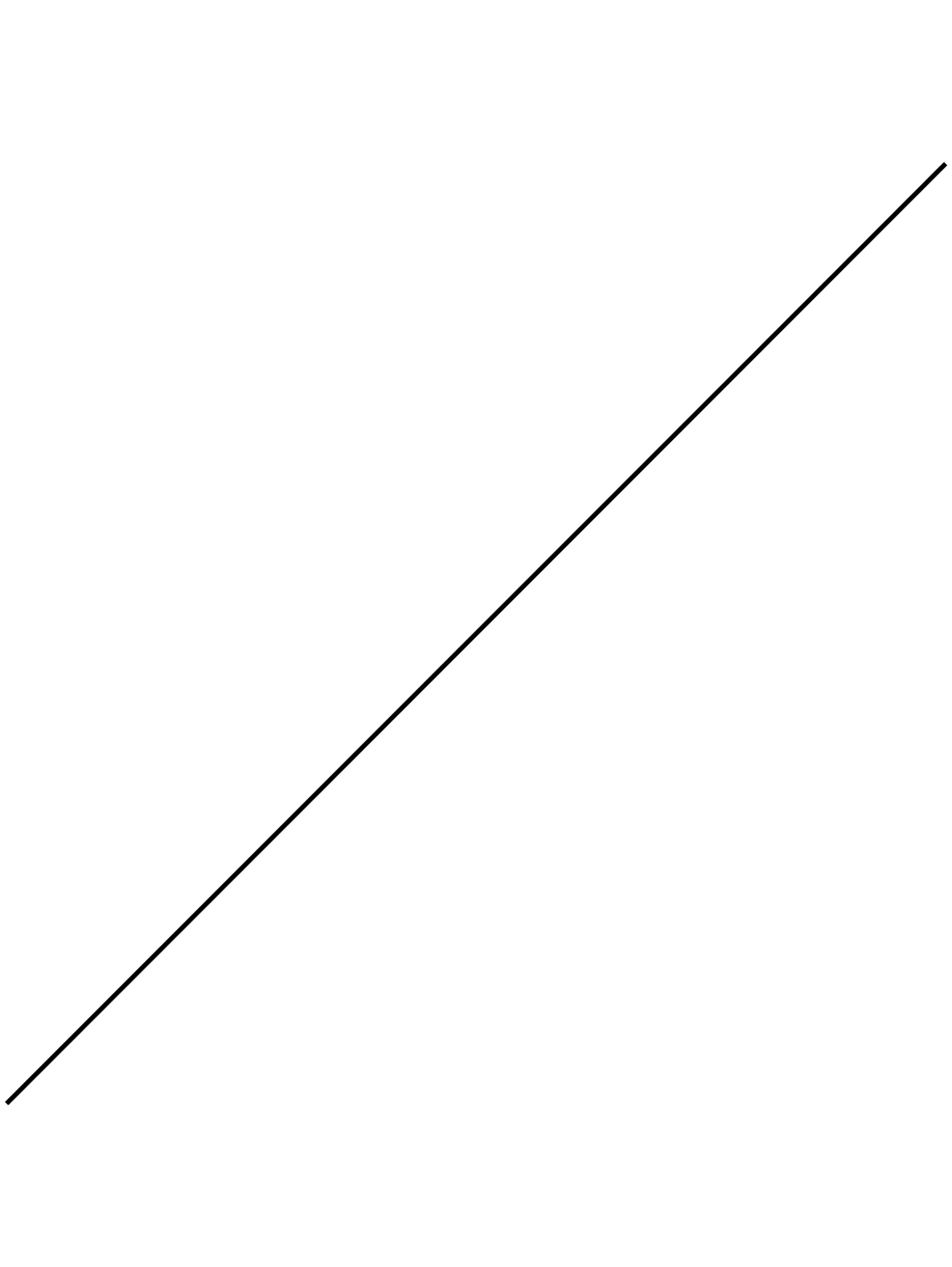} &
    \includegraphics*[height=40 mm]{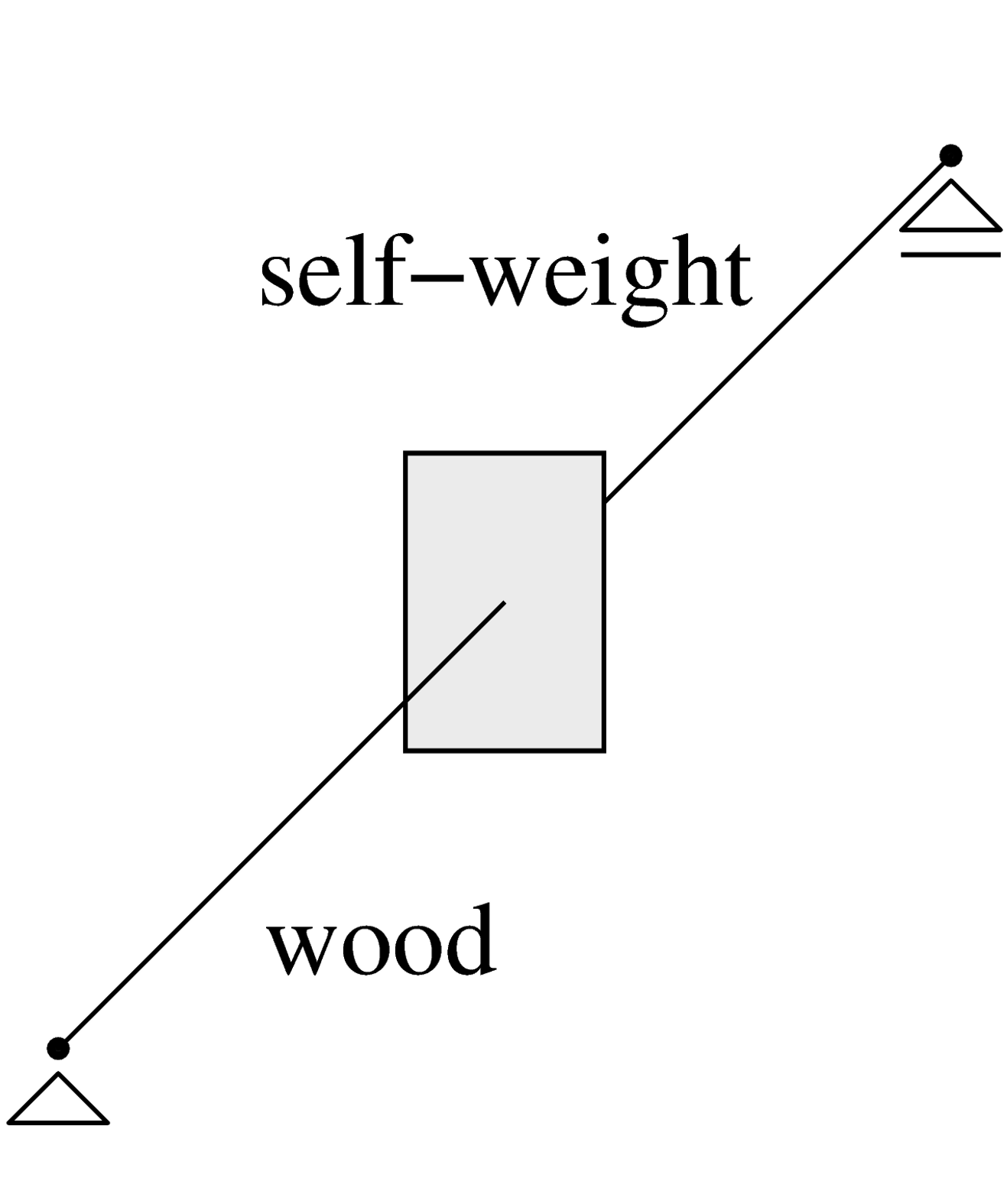} \\
    a) & b) & c)
  \end{tabular}
  \caption{Object modelling of beam with rectangular cross-section, a) architectural model, b) structural model, c) object model.}
  \label{fig:bim}
\end{figure}

The conversion can be briefly illustrated by the example of a straight beam with rectangular cross-section,~\reffig{bim}. In the usual architectural model, a beam is displayed on the output device and maintained in computer memory as a set of twelve lines topologically linked with the nodes located in eight vertices,~\reffig{bim}a. For the purpose of an effective structural analysis, this model is simplified to an one-dimensional line segment. Afterwards, the computational model is supplied with additional information, here e.g. cross-sectional characteristics, material parameters and applied loads,~\reffig{bim}c. A similar object-based approach is also applied for planar and shell entities.

\section{Software architecture}
\begin{figure}[h!]
  \centering
  \includegraphics*[]{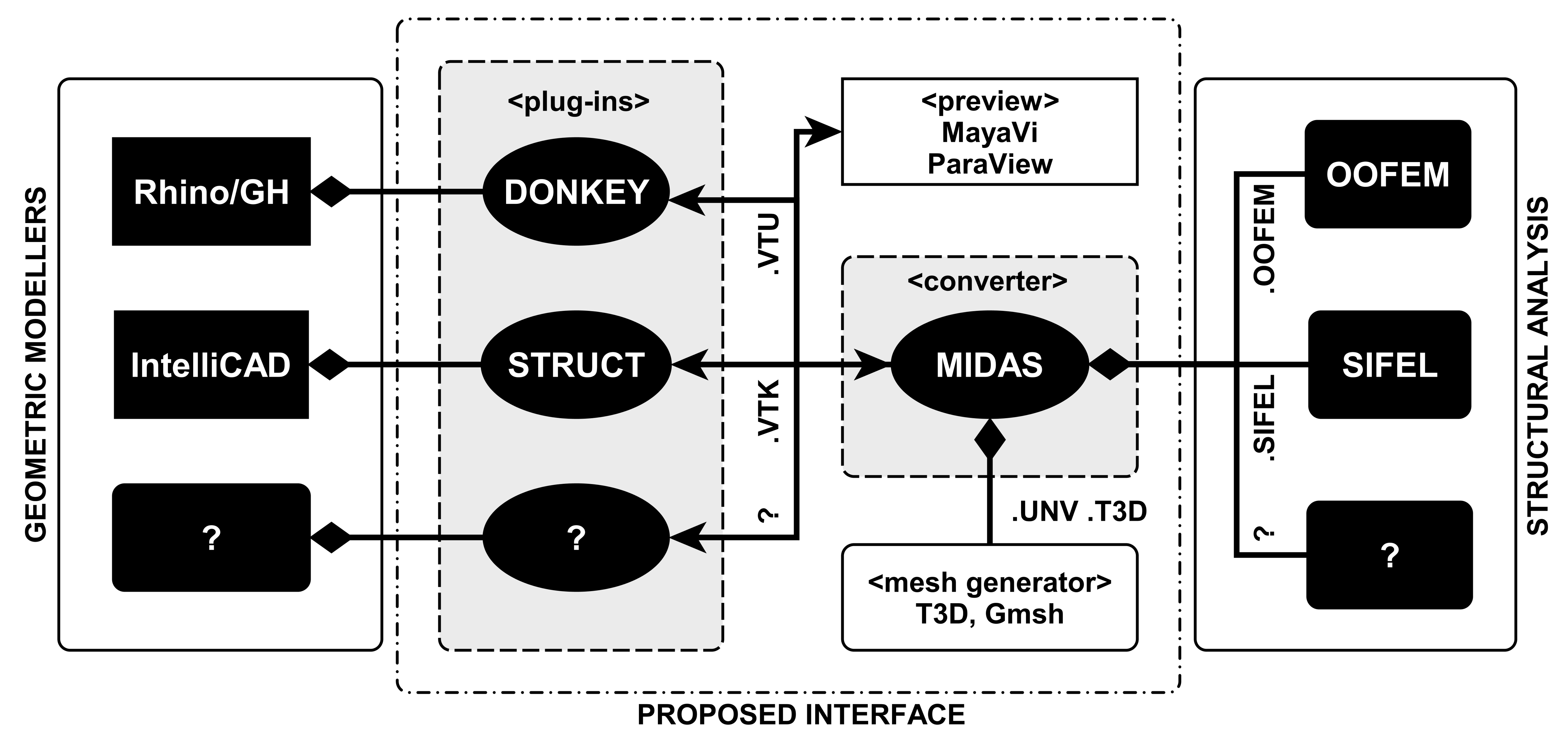}
  \caption{Flowchart of individual program components.}
  \label{fig:chain}
\end{figure}

%
The basic structure of the proposed interface is briefly outlined in~\reffig{chain}. As mentioned above, we exploit a modular approach in which each of the module is responsible for a particular action within the communication chain between structural engineers and designers. The converter and the plug-ins to geometric modellers were newly created (dashed line grey boxes in the flowchart) and released under the open source licence regulations. Existing components were used and extended when needed. Where possible, free and open source variants of particular modules (round corner boxes) were preferred for their flexibility and accessibility.

A Multifunctional Interface Between Design and Mechanical Response Solver (MIDAS)~\cite{MIDAS:www} is in the heart of the reported system. It is responsible for manipulating input and output data of structural analysis in various formats and was tested in combination with in-house developed packages OOFEM~\cite{OOFEM:homepage,OOFEMdesign}, SIFEL~\cite{SIFELhp} and proprietary system ANSYS. As for the input data, there exist several ways of generating structural models. For instance, simple benchmarks can be written directly in a text editor. On the contrary, unique models are best to be prepared by single-purpose generators, see~\refsec{GDF}. In most cases, however, the designer is expected to come in a close contact only with his favourite modeller and the corresponding plug-in, e.g. DONKEY~\cite{DONKEYhp,Kurilla2012_W4} and STRUCT~\cite{STRUCThp}. The remaining process is assumed as an automated black-box tool. The plug-in should help the user to create a structural model and provide it with additional information to run the analysis, see~\refsec{object}. The sequence of individual routines is as follows:
\begin{enumerate}
  \setlength{\parsep}{-1pt}
  \item user: architectural/geometric modeller - generation of structure's geometry;
  \item plug-in: completion of object model; forward VTK export;
  \item MIDAS: data modification and consistency check; generation of finite element (FE) package input file;
  \item FE package: structural analysis;
  \item MIDAS: output data processing; backward VTK export for visualisation purposes;
  \item plug-in: visualisation of results.
\end{enumerate}

As discussed above, typical user is expected to have only a minor experience with theoretical and computational aspects of structural analysis and the software tool should provide him with an interactive learning interface. To guarantee this, the plug-in(s) should meet the following criteria:
\begin{enumerate}
  \setlength{\parsep}{-1pt}
  \item only basic structural analysis (linear statics with truss, beam and shell elements);
  \item geometric model is clean, no confusing details are contained;
  \item material characteristics and boundary conditions can be set up in a simplified and extended regime (e.g. predefined or custom materials);
  \item interpretation of mechanical response with optional level of detail that enables designers to choose a post-processing mode
        adequate to their particular needs and knowledge (e.g. ``yes-no'' binary markers, cross-section resistance ratio
        or a full set of internal forces, displacements, strains and stresses);
  \item interactive and intuitive handling.
\end{enumerate}

The flow of data proceeds through the utility chain by means of files in various formats, see~\refsec{MIDAS}, since particular modules support different input/output. The ASCII VTK (Visualization Tool Kit)\footnote{www.vtk.org/VTK/img/file-formats.pdf} has been chosen as the primary format. It has a human readable syntax and can be visualised directly in the modeller or free visualisation tool-kits such as Paraview or MayaVi\footnote{www.paraview.org, mayavi.sourceforge.net}. Thanks to this, the data exchange can be simply controlled at any stage of software development and debugging.

Regarding the FE discretization of geometric models, we have explored two equivalent methods. Namely, the modeller triangulation toolkit, originally involved for rendering visualisation purposes, and an external mesh generator that is called from MIDAS. Since the modern architectural models mostly consist of NURBS (Non-Uniform Rational B-Spline) entities native for Rhinoceros~\cite{rhino:homepage}, the same geometry definitions are also used for mesh generation. However, this together with built-in generator sometimes leads to poor mesh quality. Thus, a more flexible way appears to consist from passing the solid geometry to MIDAS and generate the mesh by an external utility, e.g. T3D~\cite{T3d} or Gmsh~\cite{NME:NME2579,GMSH}.

\section{Prototype implementation}
%
The efficient basis of a implementation is composed of MIDAS and other two in-house developed software packages OOFEM and T3D. OOFEM is a modular finite element code for solving problems of solid, transport and fluid mechanics. T3D is a mesh generator operating on complex two- and three-dimensional domains. Both OOFEM and T3D are compiled in a minimum required configuration as dynamic libraries and linked with MIDAS. The result is released as the open source software operating on various platforms.

\subsection{MIDAS}\label{sec:MIDAS}
%
The module MIDAS~\cite{MIDAS:www} is a tool without graphical user interface designated for manipulating both input and output data of structural analysis. MIDAS's source code, written in C++, is released under GPLv3+\footnote{GPLv3+: GNU GPL version 3 or later, http://gnu.org/licenses/gpl.html} license regulations. It can work with data files of different formats - UNV, VTK, VTK XML as well as OOFEM, SIFEL, T3D and ANSYS native formats.

Recall that the input geometric model as a whole or its part can be defined by a solid geometry or a FE mesh. In the case of pure geometry, the model is discretized by T3D called from MIDAS. However, most of the subsequently listed features may be applied to both representations.

The raw data loaded by MIDAS are parsed in order to build an internal object structure representing the analysed model. On top of that, the complete topological connectivity of the model is internally assembled in such a way that each geometric element (point, edge, face, cell) is aware of other elements with shared vertices. The structured data can be analysed, modified or refined in various ways, all done by intrinsic MIDAS's features. These are, for instance, the mesh quality control, searching and merging identical nodes and finite elements, detection and removal of elements of zero dimensions, localisation and elimination of domains separated from the main body, detection of unsupported nodes of local kinematic mechanisms, parallel computing support, etc. Multiple independent non-conforming meshes can be connected utilising hanging nodes or rigid arms, thus for instance, the effect of reinforcement bars can be integrated in parent meshes. Moreover, eccentric joints of beam elements are also allowed through rigid arms, where the perpendicular distances between the beam axes are found either automatically or fed from the input.

Structural analysis output data are adjusted to conform with post-processing and visualisation.
In particular, we plot the cross-section resistance ratio $u_{el}$ that ranges from 0 to $\infty$ and has the elastic-plastic threshold $u_{el,lim}=1$.
Its evaluation is based on the {\it von Mises} yield criterion
\begin{equation}
  f \left( \sigma , k \right) = \sqrt{J_2} - k = 0.
\end{equation}
Assuming the equivalent stress in the form
\begin{eqnarray}
  & \sigma_{eq} = \sqrt{3J_2} , 
\end{eqnarray}
we can write
\begin{equation}
  u_{el} = \sigma_{eq} / R_y ,
\end{equation}
where $R_y$ denotes the yield stress and $J_2$ is the second invariant of the stress deviator~\cite{Bittnar:1996:NMS}. It is obvious that $u_{el}<1$ indicates beams loaded in elastic regime and values greater then 1.0 those witch are developing inadmissible plastic zones. We are fully aware of this indicator being rather provisional, especially for materials of anisotropic strength, however, it provides us with an instant and sufficient information on the overall stress distribution in the entire model.

Due to the license regulations covering the source code, MIDAS can be easily extended according to additional needs, e.g. when solving non-standard problems with complex geometry and topology,~\refsec{GDF}. MIDAS is the principal ingredient of the proposed methodology, as it integrates all the remaining components together. It is a surrogate for the structural engineer's expertise, thereby allowing to reduce his/her personal involvement with a post-processing kit.

In an ideal situation, architectural and structural models are identical and MIDAS converts the data received from the modeller directly to the FE solver. In particular, it selects material characteristics from the database, assigns them to the finite elements, prescribes required loads, kinematic constraints and produces the OOFEM input file. In more complicated situations, the model can be refined by making use of any of the MIDAS's features mentioned above.

\subsection{DONKEY}
%

\begin{figure}[t!]
  \centering
  \includegraphics*[]{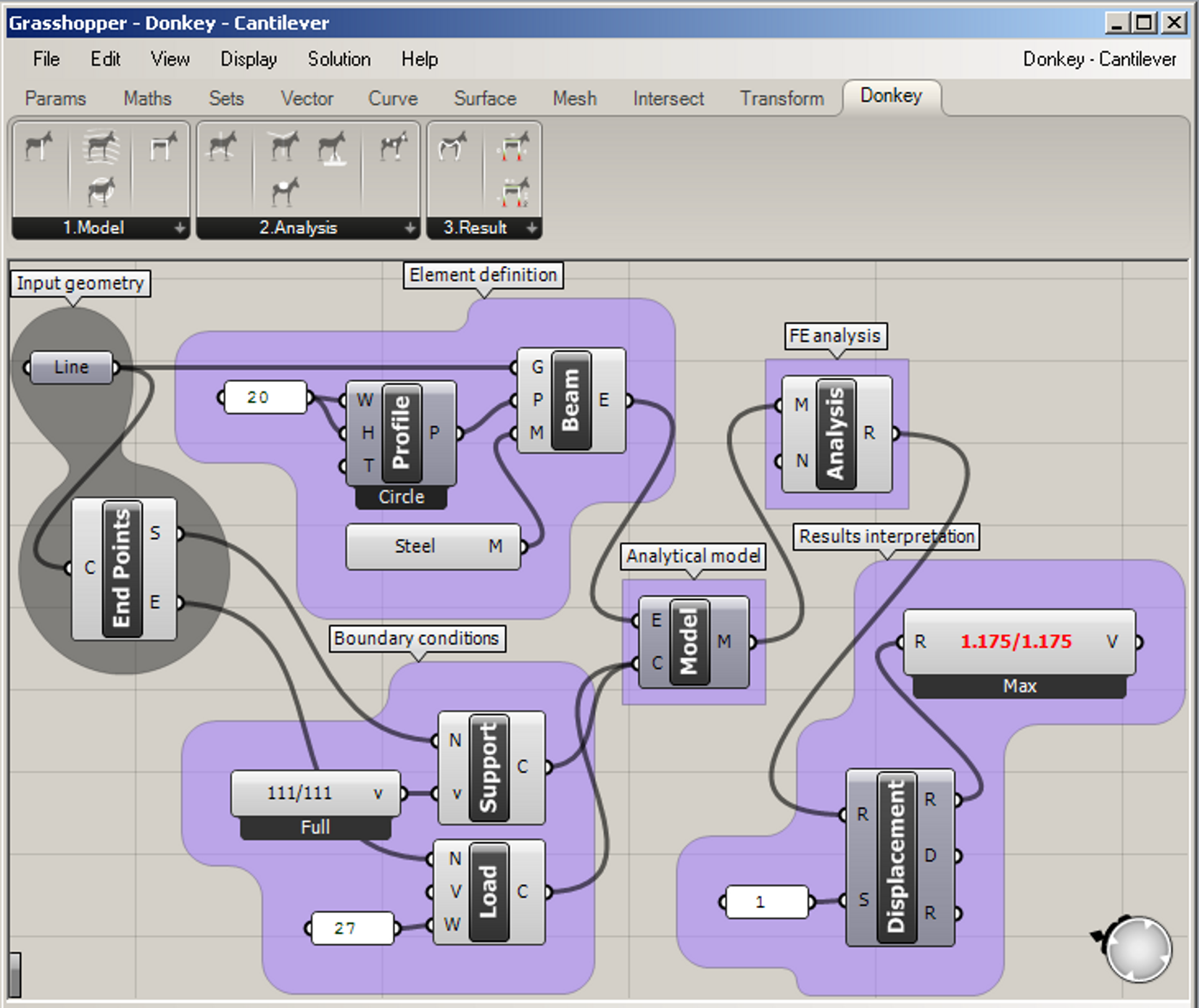}
  \caption{Demonstration visual program with DONKEY components.}
  \label{fig:donkey}
\end{figure}

The graphical algorithm editor Grasshopper~\cite{grasshopper:homepage}, closely integrated with the NURBS-based 3D modelling tool Rhinoceros, was chosen as the coding framework of the plug-in DONKEY~\cite{DONKEYhp,Kurilla2012_W4}. Grasshopper is a visual programming tool for procedural modelling popular among academics and professionals. It allows designers to generate simple geometries as easily as the awe-inspiring ones, still preserving possibility of interactive modifications. Programs are created by dragging components with particular functionality onto a canvas. The outputs of these components are then connected to  inputs of subsequent components. In this environment, DONKEY is accessible as a set of components in a separate tab of Grasshopper's menu, see~\reffig{donkey}. Properties of DONKEY components highlighted in~\reffig{donkey} are demonstrated on the example of a cantilever of $1000$\,mm in length and circular cross-section of $20$\,mm in diameter, being subject to the vertical force of $264$\,N, corresponding to $27$\,kg, acting at the unconstrained tip.

\begin{figure}[t!]
  \centering
  \includegraphics*[width=120mm]{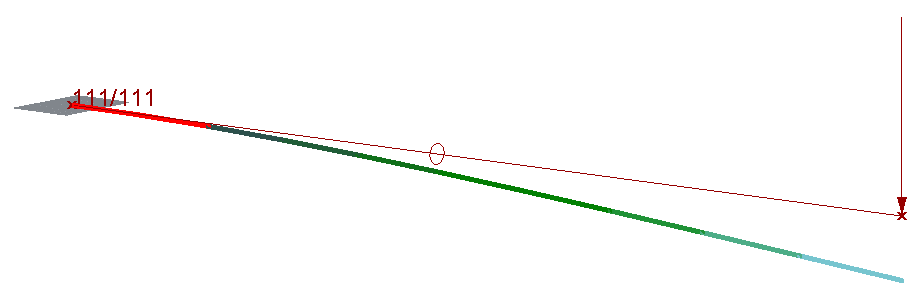}
  \caption{Cantilever and cross-section resistance ratio drawn on deformed shape.}
  \label{fig:donkey-rhino}
\end{figure}

%
In the first step, a user creates a geometric model, appropriate for structural analysis, by using a fully automated tool (algorithmic architecture) or a standard drawing procedure (human input based CAD layout). In this particular example, a single line and its end points are obtained by Grasshoppers' built-in functions {\sc Line} and {\sc End Points}, notice the group {\sc Geometric model} in~\reffig{donkey}. Within the second step, each of the entities is provided with the information necessary for numerical analysis, thereby the solid geometry becoming structural model. In particular, circular cross-section and steel material was assigned to the line by components {\sc Profile} and {\sc Steel}. Next, a constraint (all displacements and rotational degrees of freedom constrained by default) is applied to one, component {\sc Support}, and a force load to the opposite end point of the beam, component {\sc Load}. The model, component {\sc Model} is exported to a VTK XML file, ~\reftab{VTKtab}, and sent to MIDAS, component {\sc Analysis}. It is further discretized by calling T3D and analysed in OOFEM. Finally, the cross-section resistance ratio and mechanical quantities such as strains, stresses and displacements can be visualised by corresponding components, see the group {\sc Result interpretation}. The screen-shot of Rhinoceros view-port captures the structural model and the cross-section resistance ratio drawn on the deformed cantilever,~\reffig{donkey-rhino}. The highest calculated resistance ratio is 1.175, as visible in~\reffig{donkey}, component {\sc Max}.

\subsection{Exchange data file format}
%
The flow of data proceeds between DONKEY and MIDAS by means of files in VTK XML format,~\reftab{VTKtab}.
The geometry is defined through initial pair of data blocks proceeded by the POINTS and CELLS keywords.
Structural properties assigned to geometric elements are stored in POINT\_DATA and CELL\_DATA sections.
The unstructured section AppendedData contains generic information of the project's name, material specifications, cross-section characteristics etc.
To speed up the data flow, the ASCII is replaced with the binary format and the particular files are stored in virtual memory instead of the hard drive.

%
%
\begin{table}[ht!] \centering
  {\tableset{0}{6}
    \begin{tabular}{|m{135mm}|}
      \hline \vspace{0mm}
             {\scriptsize
\begin{verbatim}
<VTKFile type="PolyData" version="0.1" byte_order="LittleEndian">
  <PolyData>
    <Piece NumberOfPoints="2" NumberOfLines="1">
      <Points>
        <DataArray type="Float32" NumberOfComponents="3" format="ascii">
             0.0 0.0 0.0
          1000.0 0.0 0.0
        </DataArray>
      </Points>
      <Lines>
        <DataArray format="ascii" type="Int32" Name="connectivity"> 0 1 </DataArray>
        <DataArray format="ascii" type="Int32" Name="offsets"> 2 </DataArray>
      </Lines>
      <PointData>
        <DataArray format="ascii" type="Int32" Name="Boundary_Conditions" NumOfComp="6">
          1 1 1 1 1 1
          0 0 0 0 0 0
        </DataArray>
        <DataArray format="ascii" type="Int32" Name="ID_BOUNDARY_CONDITION">
          0
          1
        </DataArray>
      </PointData>
      <CellData>
        <DataArray format="ascii" type="Int32" Name="ID_CROSS-SECTION"> 2 </DataArray>
        <DataArray format="ascii" type="Int32" Name="ID_MATERIAL"> 1 </DataArray>
      </CellData>
    </Piece>
  </PolyData>
  <AppendedData>
    _
    <Characteristics>
      <COMMENT> <item> example - cantilever </item> </COMMENT>
      <CROSS-SECTIONS Number="2">
        <item> 1 Rectangle width 0.1 height 0.2 refNode y -2 </item>
        <item> 2 Circle width 20.0 </item>
      </CROSS-SECTIONS>
      <MATERIALS Number="1">
        <item> 1 IsoLinEl  E 210.0e+03  nu 0.20  tAlpha 0.000012  density 7850.0e-09 </item>
      </MATERIALS>
      <BOUNDARY_CONDITIONS Number="1">
        <item> 1 NodalLoad components 6 -264.777 0.0 0.0  0.0 0.0 0.0 </item>
      </BOUNDARY_CONDITIONS>
    </Characteristics>
  </AppendedData>
</VTKFile>
\end{verbatim}
      }
      \\[0mm] \hline
    \end{tabular}
  }
  \caption{VTK XML file generated by Donkey.}
  \label{tab:VTKtab}
\end{table}
%

\section{Case studies}
%
The proposed concept of integrated design is illustrated on three case studies. These were carried out in a close collaboration involving architectural studio FLOW at Faculty of Architecture in CTU in Prague (FA CTU), see~\cite{Florian:stud,Florian:arch}, CUBESPACE studio and a free-lance artist \mbox{Federico} \mbox{D\'{i}az}\footnote{www.cubespace.eu, www.fediaz.com}. All the contributors are engaged in algorithmic architecture featuring complex forms generated by computer algorithms that are driven by human-entered aesthetic and functional contexts~\cite{DigCit}. Since these structures differ from traditional ones, it is difficult to reliably assess their mechanical behaviour without the computer-aided structural analyses. However, a detailed simulation of their response would be too prohibitive, considering rather early phases of the projects. The results represent the final responses of manually (The Leonardo Bridge, Annelida) and automatically optimised structures (GDF).

The first of three case studies, Kurilla's Annelida bridge~\cite{Kurilla:arch}, represents a  heterogeneous geometry composed of shells and girders, which requires a significant reduction to become an acceptable structural model. On the contrary, the self-supporting Leonardo's bridge is much less complicated and the architectural wire model almost coincides with that for structural analysis.
Finally, we demonstrate the full power of MIDAS interface on the investigation to very complex sculpture, Geometric Death Frequency-141, by Federico D\'{i}az~\cite{GDF:book}.


\subsection{Annelida}
\begin{figure}[t!]
  \centering
  \begin{tabular}{c@{\hspace{4mm}}c}
    \includegraphics*[width=73 mm]{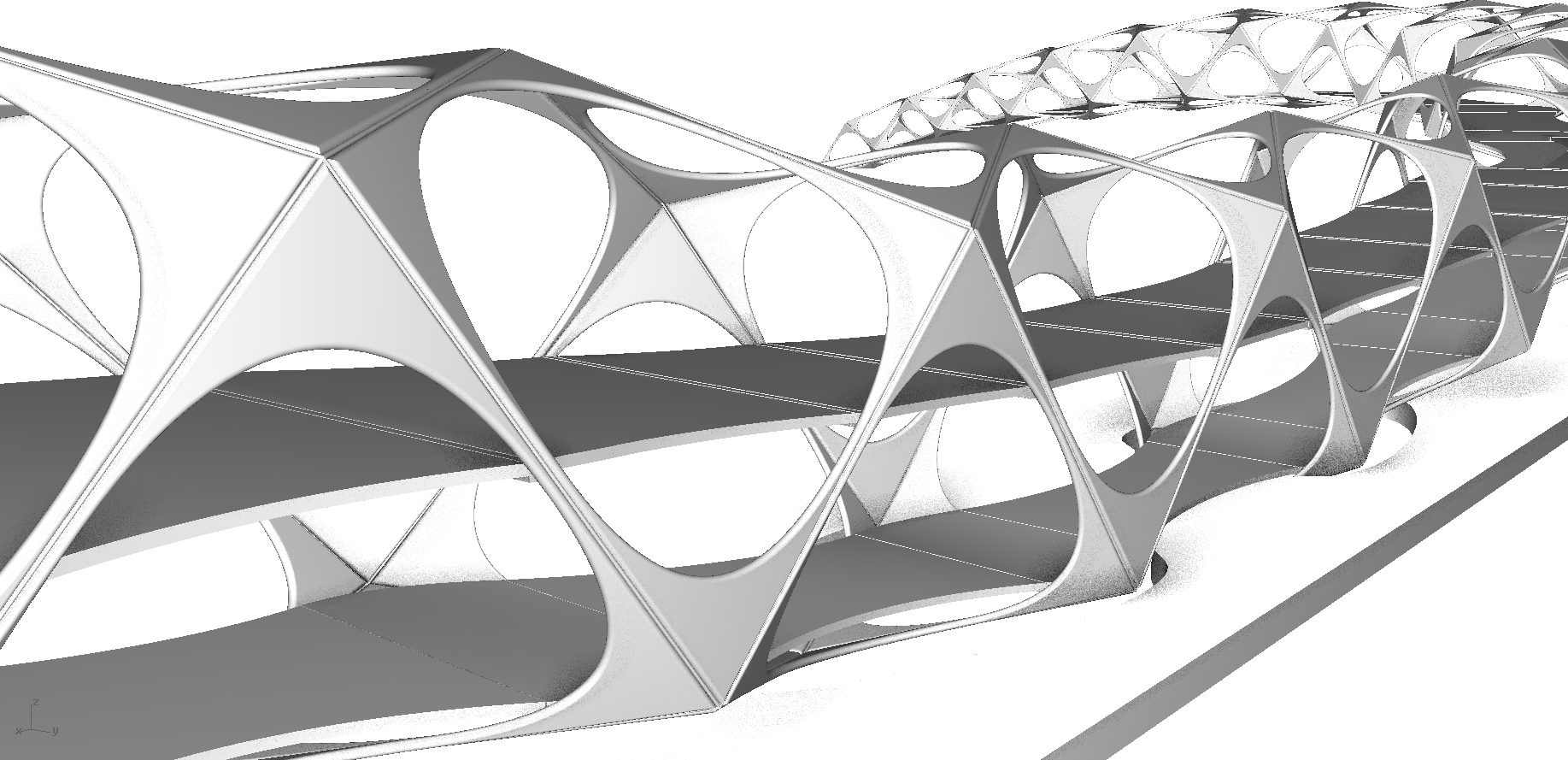} &
    \includegraphics*[width=65 mm]{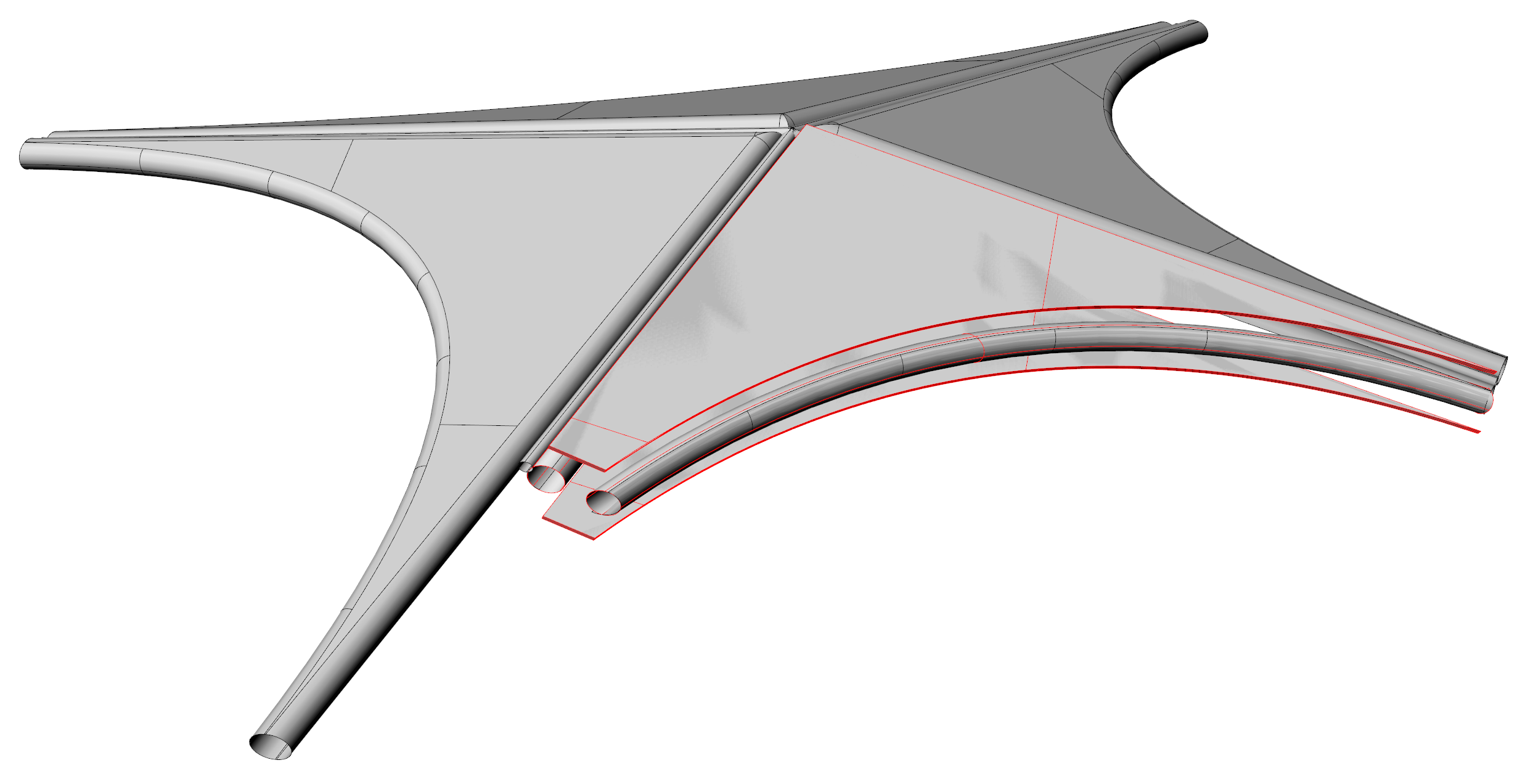} \\
    a) & b) \\
    \includegraphics*[width=73 mm]{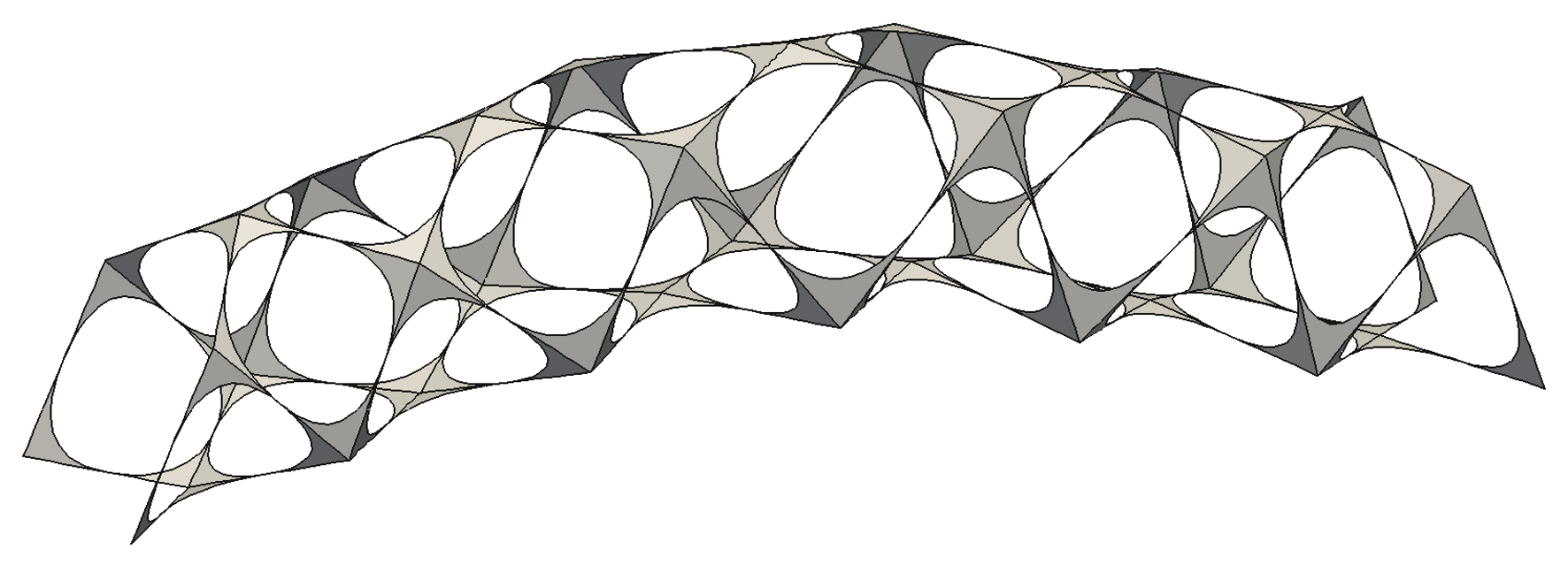} &
    \includegraphics*[width=65 mm]{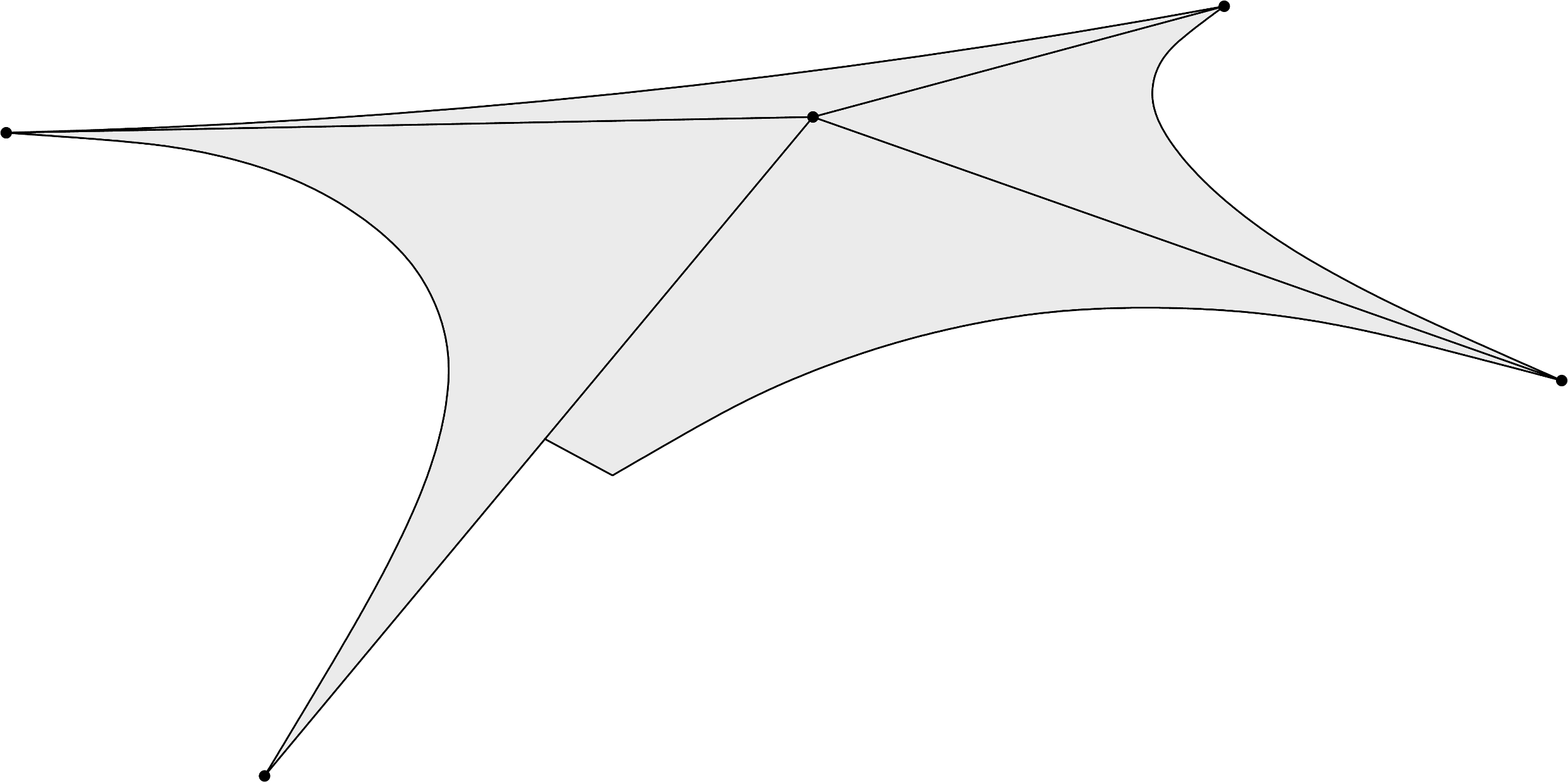} \\
    c) & d)
  \end{tabular}
  \caption{Annelida, a) complex architectural model, b) detail of joint,
                     c) structural model,            d) structural model of joint.}
  \label{fig:annelida}
\end{figure}

%
Annelida bridge exemplifies a complex task whose computational model has to be significantly simplified before the structural analysis execution. The bridge, made up of steel as suggested by  Luk\'{a}\v{s} Kurilla\footnote{http://www.studioflorian.com/projekty/63-lukas-kurilla-annelida}~\cite{Kurilla:arch}, was truncated for demonstrative purposes to a $44\times8\times12$\,m segment. The frame is composed of straight and arc tubes of circular cross-sections that form a repeating geometric pattern of distorted rectangles with circular openings,~\reffig{annelida}a. The frame vertices are reinforced with a pair of concave steel plates of mutual distance equal to the outer diameter of the frame tube being aligned with,~\reffig{annelida}b. The structure is supported at two pairs of points, each pair located at a single bridge end. The parametrized model was generated automatically by means of a single purpose script. These parameters were optimised on the basis of the resulting mechanical response.

As shown in~\reffig{annelida}b, the architectural model has been created including all the details specific to design features and omitting any computational simplifications. Even with a high performance computer at hand, it would be barely possible to generate a mesh of shell or volume finite elements resolving the model in detail, see e.g.~\cite{SandS}, as the elements in the tube walls would be much smaller then those in reinforcing plates. Such a fine FE discretization would result in an excessive computational overhead. Furthermore, technicalities, such as connecting the pairs of straight frame bars would be difficult within the ``detailed discretization'' concept as well, since these are slightly non-parallel thanks to the distorted topology of the entire structure. For these reasons the script was modified to generate a simplified architectural model where the frame bars reduce to the centroid axes of zero cross-sectional area and only a single mid surface represents the twin corner haunches,~\reffig{annelida}d.

Next steps were identical to the previous cantilever example, with the exception of FE mesh generation that was executed in Rhinoceros. Identical coordinates were prescribed to all nodes at the contact among beam and shell elements and arising multiplicities were merged in MIDAS.

\subsection{The Leonardo Bridge}
\begin{figure}[t]
  \centering
  \includegraphics*[]{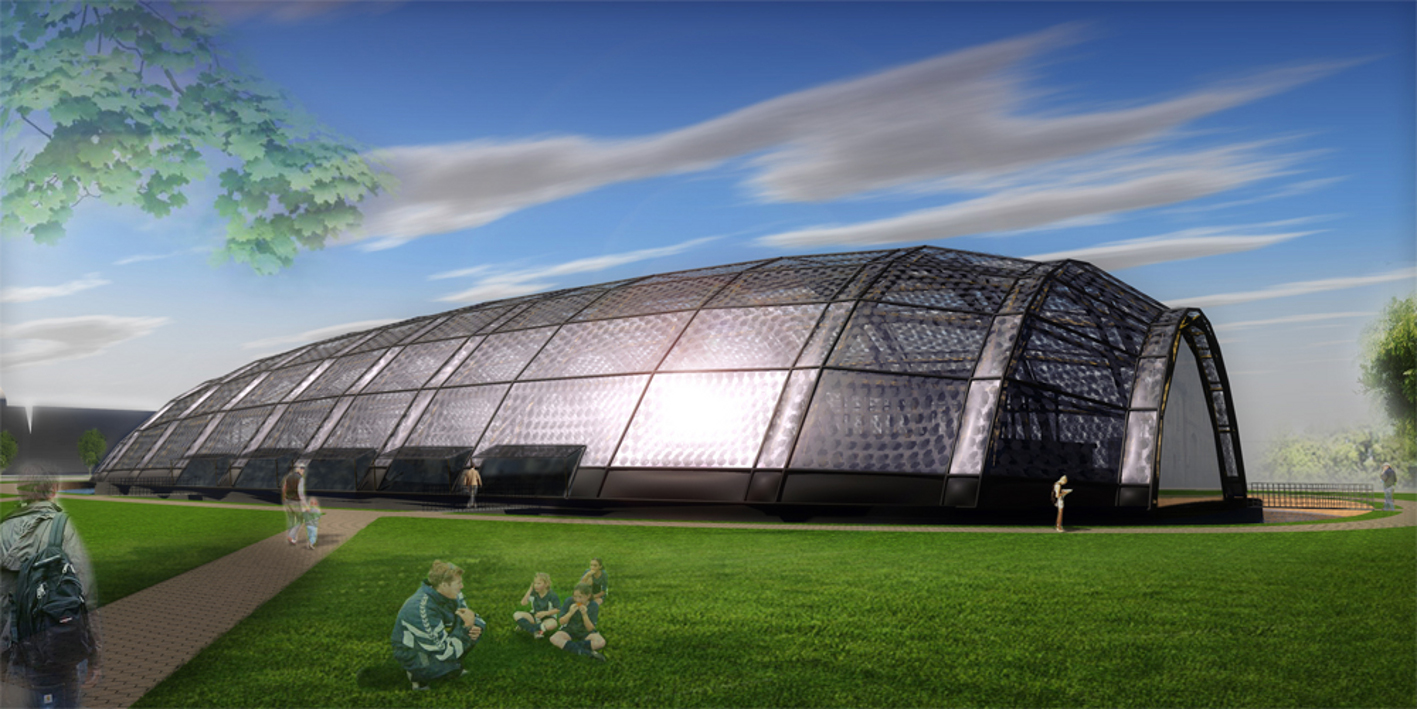} \\
  \caption{Sports hall.}
  \label{fig:leon-hala}
\end{figure}

\begin{figure}[t]
  \centering
  \begin{tabular}{c@{\hspace{4mm}}c}
    \includegraphics*[height=42 mm]{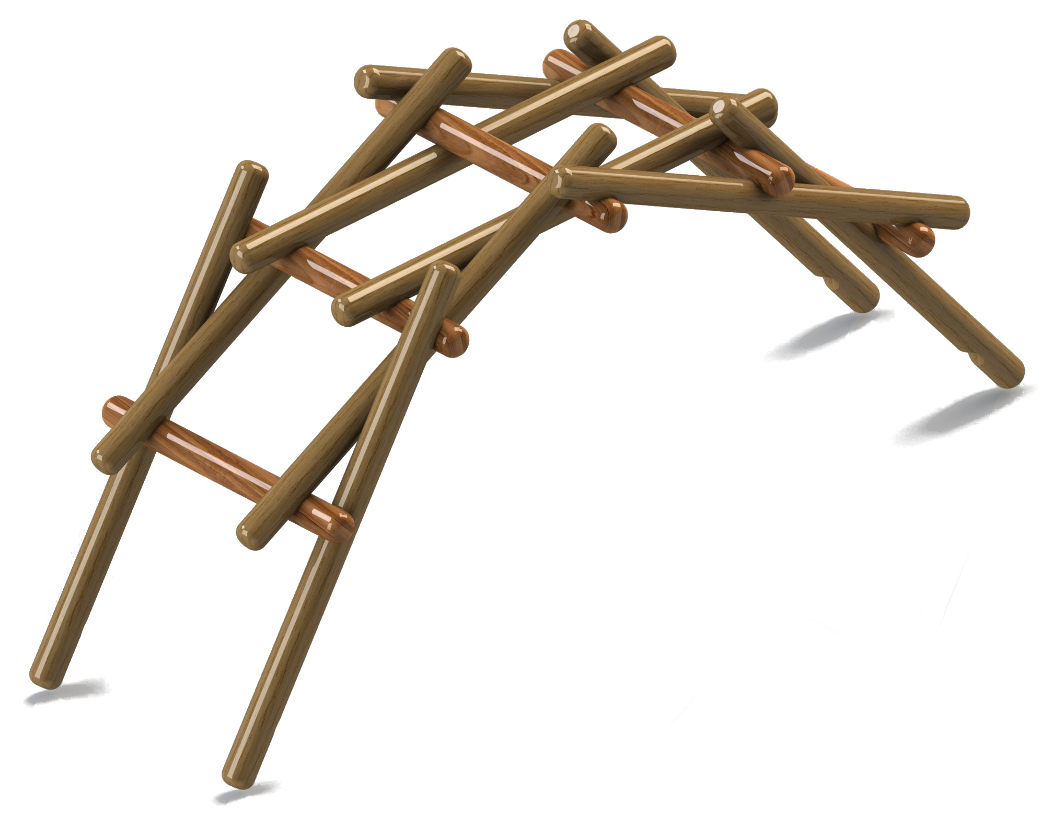} &
    \includegraphics*[height=42 mm]{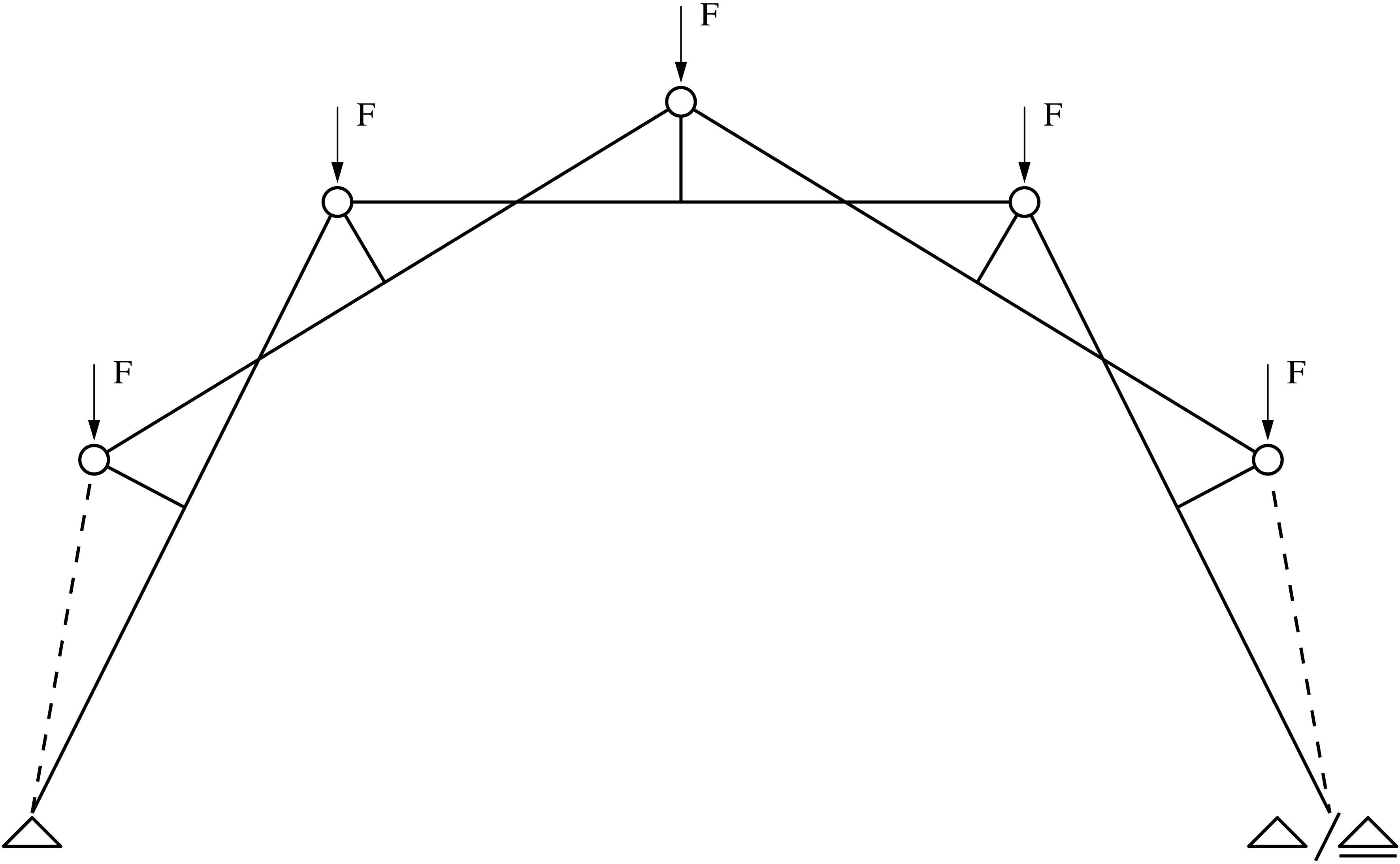}   \\
    a) & b)
  \end{tabular}
  \caption{ Self-supporting arch a) 3d scheme, b) 2d structural model. Dashed lines indicate {\it closed} (with) and {\it open} (without) variant.}
  \label{fig:leon-concept}
\end{figure}

%
By means of the Leonardo bridge, we would like to demonstrate learn and form-finding capabilities of the proposed software. The project of a sports hall\footnote{www.studioflorian.com/projekty/184-martin-cisar-mestska-sportovni-hala-v-kutne-hore} for up to 300 spectators was designed by Martin C\'{\i}sa\v{r},~\reffig{leon-hala}, and analysed by Karol\'{\i}na Ma\v{s}kov\'{a}~\cite{krolina}, the undergraduate students at FA CTU and Faculty of Civil Engineering of CTU in Prague, respectively. The structure is composed of fourteen arch sections inspired by Leonardo da Vinci's self-supporting bridge, famous for its ingenious simplicity,~\reffig{leon-concept}a. Besides its structural efficiency, the system is known for the self-locking joints, which enable fast erection without fasteners and easy disassembly.

\begin{figure}[t]
  \centering
  \begin{tabular}{c@{\hspace{4mm}}l}
    a)  &  \includegraphics*[]{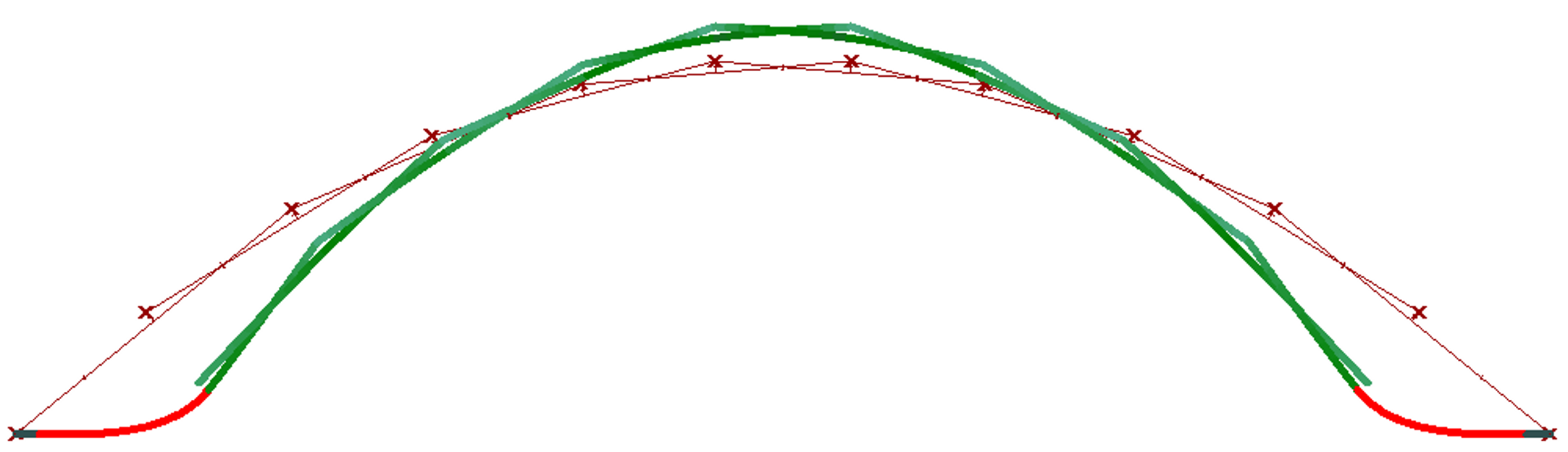}  \\
    b)  &  \includegraphics*[]{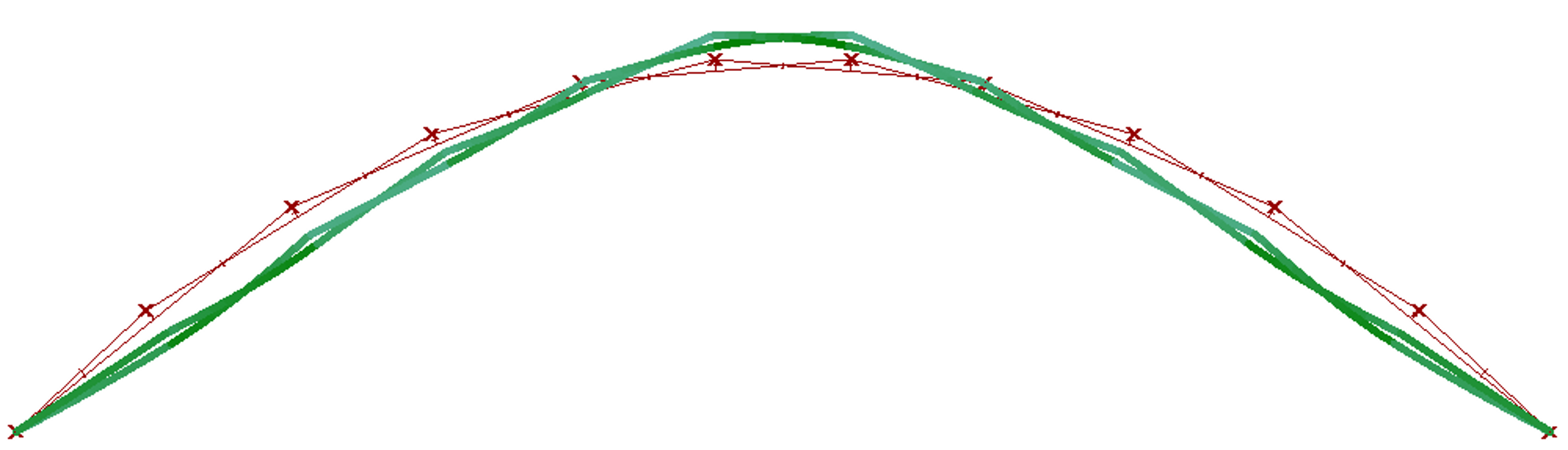}  \\
    c)  &  \includegraphics*[]{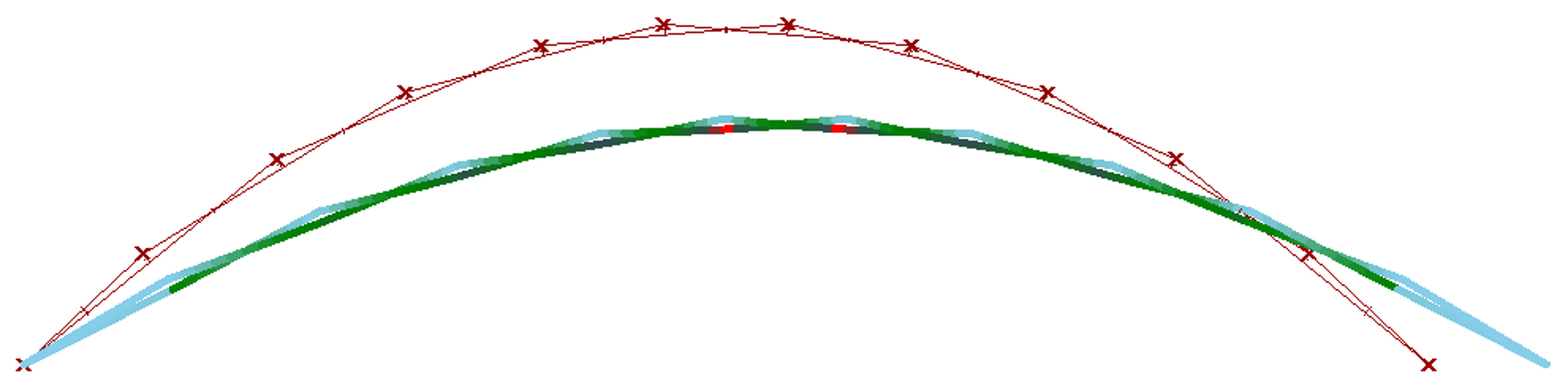}     
  \end{tabular}
  \caption{Three distinct models with identical setup and displacement -
    a) {\it open} variant, b) {\it closed} variant and c) {\it closed} variant with mobile support.}
  \label{fig:leon-vary}
\end{figure}

The typical arch is of $35$\,m in length and $13$\,m in height. It is assembled of timber beams rectangular in cross-section, which must resist loading by the self-weight (the segments themselves plus the roofs dead load) and standardised weight of snow. The structural model,~\reffig{leon-concept}b, was generated by an algorithm with parameters of the number of segments and lengths, and cross-section dimensions of individual beams. At the first instance, three distinct models with identical setup of design parameters were compared,~\reffig{leon-vary}.
Although the {\it open} variant,~\reffig{leon-concept} and~\reffig{leon-vary}a, is more advantageous from the application point of view\footnote{simple.wikipedia.org/wiki/File:Da\_vinci\_bridge.jpg, http://www.rlt.com/20101}, we observe the significant local displacements arise beams adjacent to applied supports, compared to its {\it closed} counterpart,~\reffig{leon-vary}b. In particular, the maximum total displacements and the cross-section resistance ratio are $44$\,mm and $0.45$ for the closed variant in contrast to $154$\,mm and $1.42$ for the open one. Another fundamental distinction in overall structural response brings the removal of horizontal constraint in one of the supports,~\reffig{leon-concept}b. Besides the different shape of flexural curve,~\reffig{leon-vary}b and~\reffig{leon-vary}c, it is obvious that the self locking mechanism is fully allowed only for the arch with the mobile support due to the negative bending moment at the top of the arch in~\reffig{leon-vary}b. 

\begin{figure}[t!]
  \centering
  \begin{tabular}{c@{\hspace{1mm}}c@{\hspace{1mm}}c}
    \includegraphics*[width=45 mm]{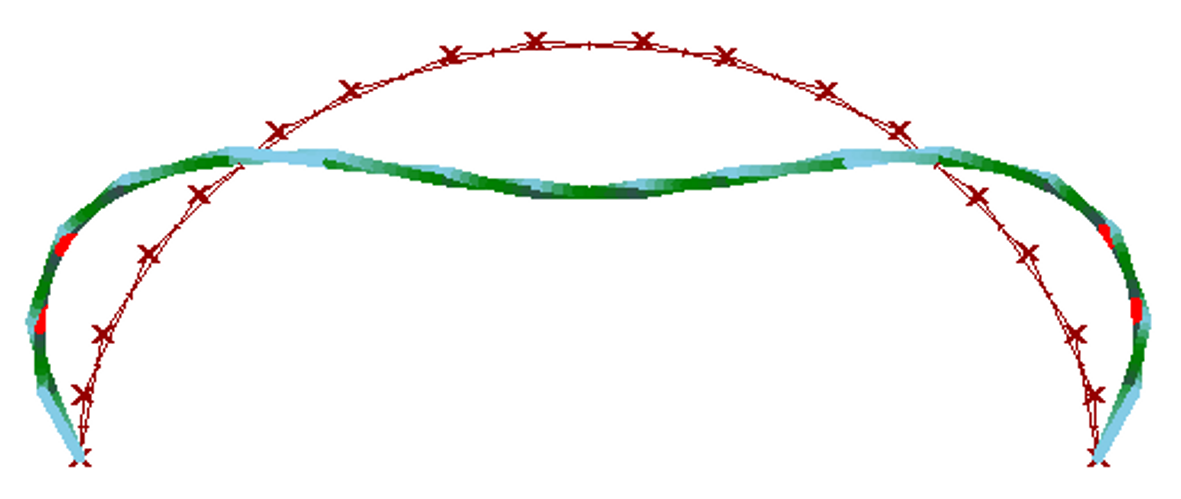} &
    \includegraphics*[width=45 mm]{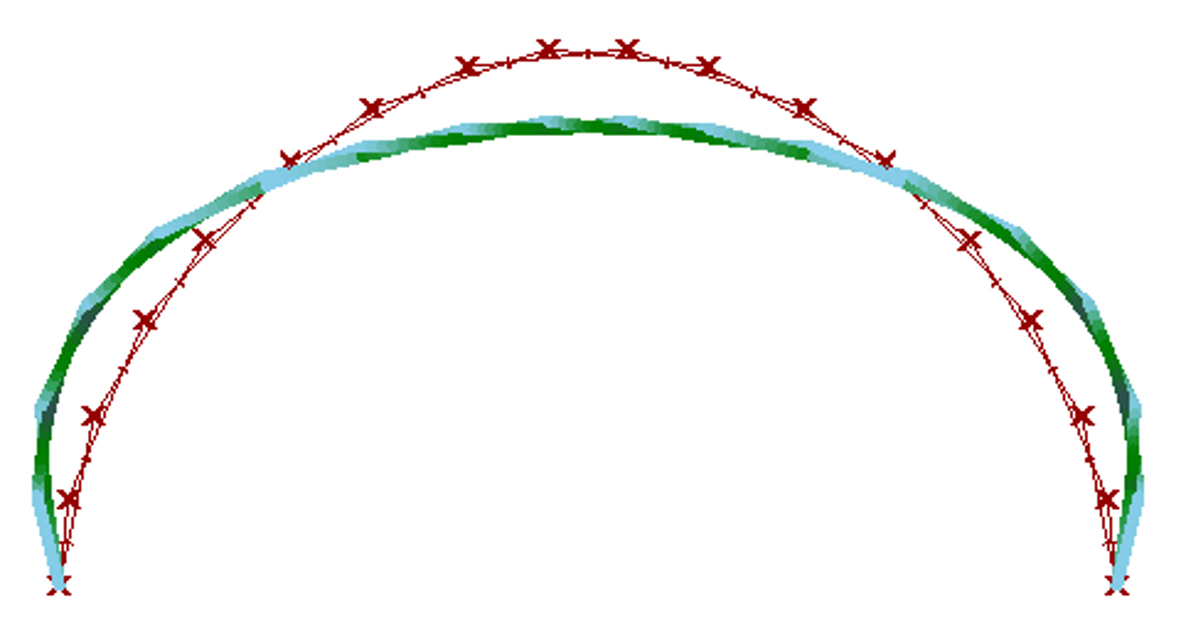} &
    \includegraphics*[width=45 mm]{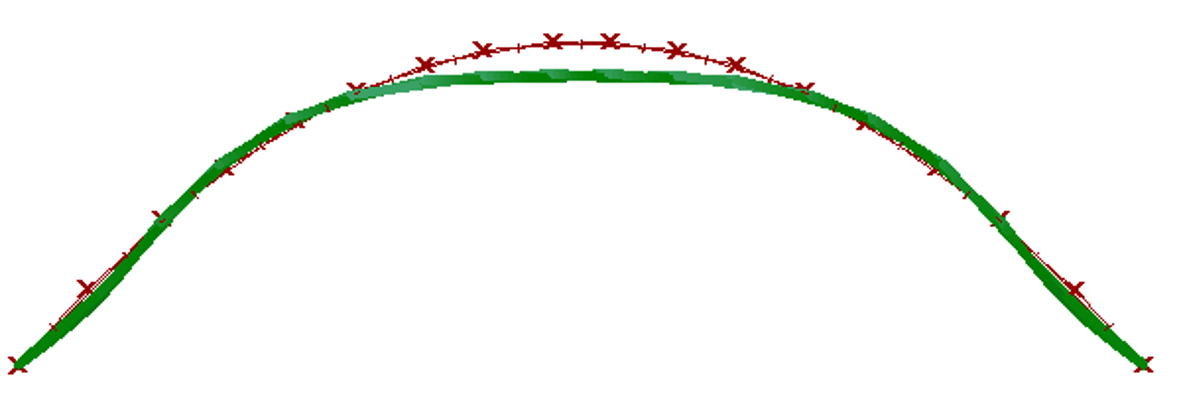} \\
  \end{tabular}
  \caption{Examples of various shape variants, undeformed and deformed shapes.}
  \label{fig:leon-optim}
\end{figure}

Resulting from the optimisation process above, the closed-form Leonardo scheme with fixed horizontal degrees of freedom was selected for the subsequent intuitive form-finding process shown in~\reffig{leon-optim}. The arch shape and cross-section dimensions were adjusted by making use of a parametric script in order to minimise displacements and cross-section resistance ratio of beam elements, yielding to the optimal shape. 

\subsection{Geometric Death Frequency-141}\label{sec:GDF}
%
The last example is to demonstrate the MIDAS's capability in application to a geometrically complex artwork with a cellular substructure made up of synthetic materials, the Geometric death Frequency-141\footnote{See www.massmoca.org/event\_details.php?id=549 for more details; the fabrication process can be found at http://vimeo.com/16019145.}
(GDF), installed by \mbox{Federico} \mbox{D\'{i}az} in the exterior of MASS MoCA (Massachusetts Museum of Contemporary Art) exhibition area in 2010~\cite{GDF:book}.

\begin{figure}[h!]
  \centering
  \includegraphics*[]{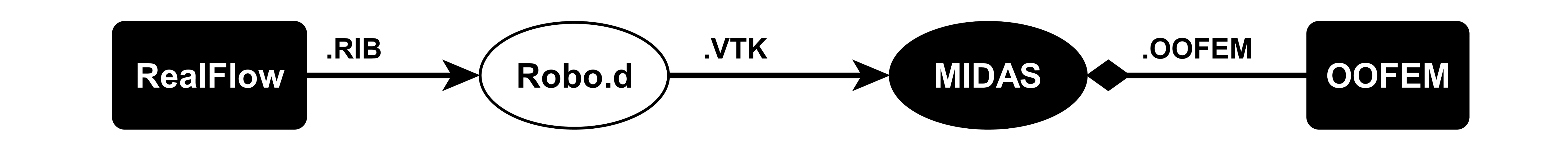}
  \caption{GDF - flowchart of individual program components.}
  \label{fig:GDF:modules}
\end{figure}

\begin{figure}[t!]
  \centering
  
  \includegraphics*[width=120mm]{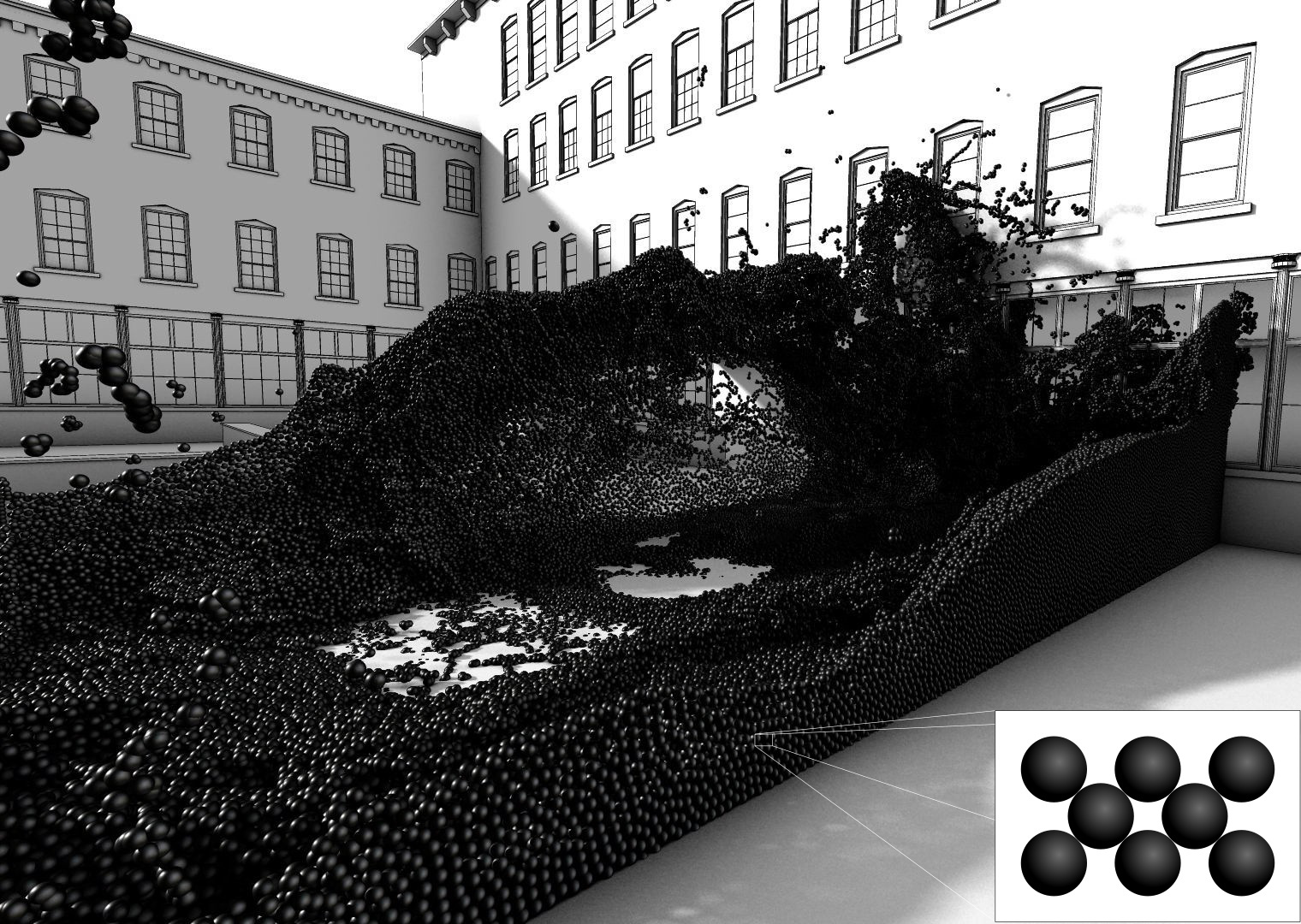}
  \caption{Geometric Death Frequency-141, visualisation.}
  \label{fig:GDF:model}
  
  \includegraphics*[width=120mm]{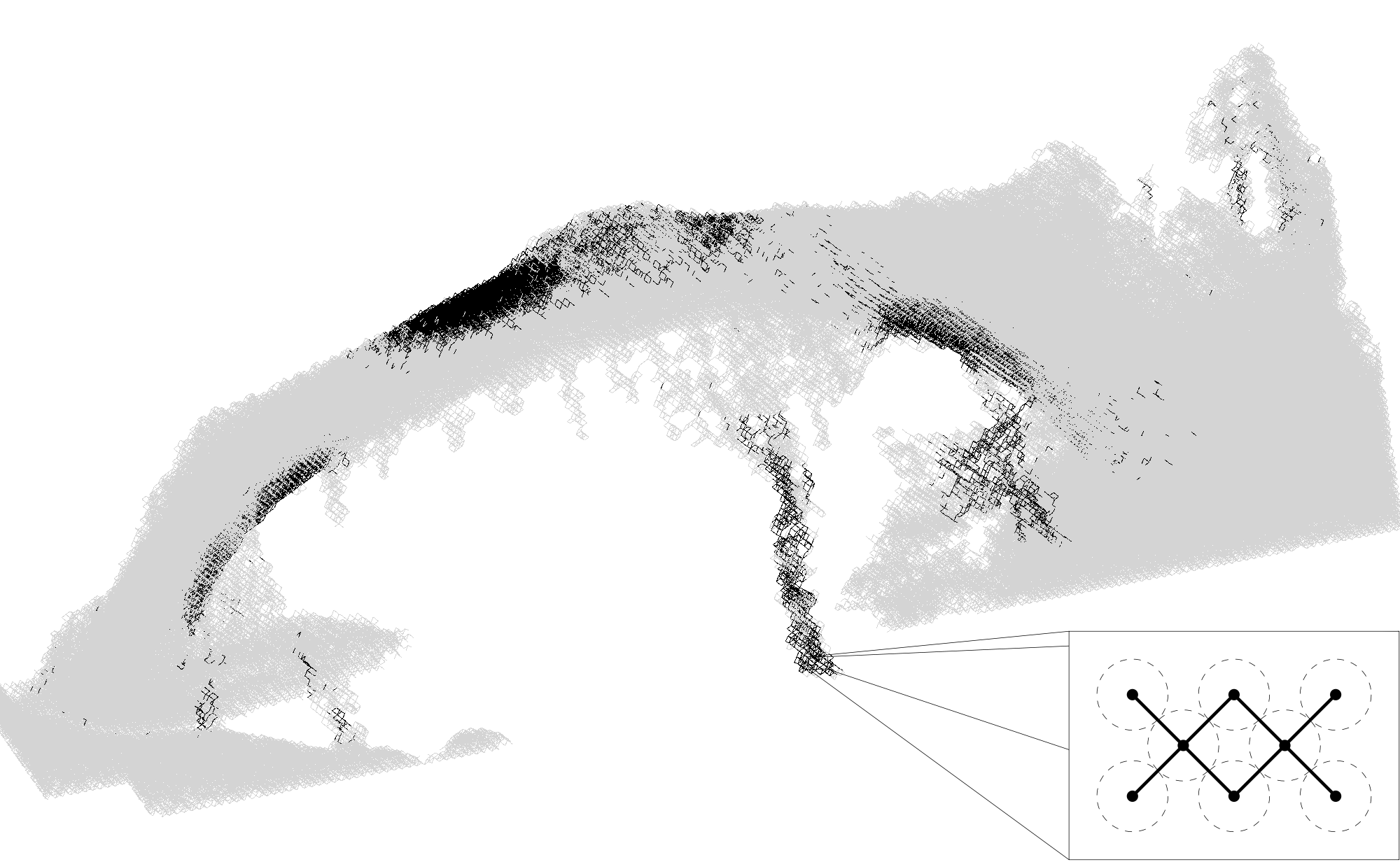}
  \caption{Arch-shaped part of GDF - cross-section resistance ratio rendered as bi-coloured scheme. Black elements indicate values greater than 1.0.}
  \label{fig:GDF:mesh}
\end{figure}

GDF represents the 141-st frame of the fluid flow analysis of a certain amount of liquid suddenly entering a closed box. The fluid motion was simulated numerically by RealFlow~\cite{RealFlow} and the particular frame was selected as the starting point for the subsequent optimisation process based on the static response. The emerging wave-like form was spatially filled up with hollow Acrylonitrile Butadiene Styrene (ABS) balls of $47$\,mm in diameter and $1$\,mm wall thickness by means of single-purpose tool Robo.d~\cite{Kurilla2012_RobArch},~\reffig{GDF:modules}. Nearly 420 thousand of balls has been assembled in a regular grid and glued together in contact points, thereby forming the self-supporting structure,~\reffig{GDF:model}. The huge amount of basic spherical cells made the manual fabrication and quality control management of all contact details unfeasible. Hence the entire process has been fully robotized.

Due to GDF's structural complexity, the mechanical response to applied loads (dead load, snow weight) was difficult in a fully automatic way by making use of the basic MIDAS functionality. Moreover, it was required by the author's team to implement additional functions for a decision management based on a priory values defined by an expert. The output data were therefore simplified to bi-coloured yes-no diagrams (beams with exceeded bearing capacity are in black,~\reffig{GDF:mesh}).

To speed-up the numerical analysis, only the compact arch-shaped part of the structure comprising of about 250 thousands balls was considered as critical. The sphere-shell sponge-like composite was transformed into a beam finite element mesh with nodes placed in the sphere centres. Thus, the beam elements represent hourglass like rotational surfaces made up of two half-spheres connected at their poles by a droplet of glue,~\reffigs{GDF:model}{GDF:mesh}. Although such a geometry yields a variable stiffness, the beams were considered as prismatic and with averaged material characteristics. The bearing capacity of the homogenized beams, normal and bending stiffness were obtained experimentally by the load test of several cantilever girders consisting of ten axially aligned balls.
The measured quantities were verified by a detailed FE analysis of a three-dimensional model with the balls and glue joints precisely resolved. The material parameters of ABS plastic and the glue were provided by the manufacturer. Finally, structural supports (displacement constraints) of the arch model were applied to all nodes representing contact points among spheres and the horizontal base.

The transformation process of GDF solid representation into the FE model was controlled by Robo.d. Despite the fully automated conversion, the raw mesh was further validated by MIDAS interface. First, nodal and element duplicities were eliminated. The nodes and elements separated from the central body mass (arising from separated drops or splashes of liquid) were identified and excluded from the analysis and subsequently from the sculpture itself~\cite{GDF:book}. Loads and boundary conditions have been applied after the adjustment.

The FE model was exported to VTK file and revised visually in order to find any major defects owing to the automatic processing (model connectivity, overall geometrical deviations between FE and the solid model, etc.). Next, it was controlled once again by the OOFEM preprocessor routines and analysed. The resulting mechanical quantities were post-processed by MIDAS and visualised in Paraview,~\reffig{GDF:mesh}.

The numerical model contained about 800 thousand degrees of freedom. Therefore, the iterative IML~\cite{SPARSE} solver of the global algebraic system with incomplete Cholesky preconditioning was used. This solver, however, exhibits poor convergence for structures with non-uniform stiffness distribution. In this particular case, such an inhomogeneity was attributed to elongated splashes of the liquid (dead arms). Thus, the function eliminating the arms of 1 to 2 balls in diameter was further implemented to MIDAS. This led to removal of $1$\,\% of finite elements with a negligible effect on overall response while reducing the computational time down to fractions of the original time.

Solving the structure, certain floppy spots had been detected,~\reffig{GDF:mesh}. The shape evolution then proceeded to choosing yet another frame of the fluid stream and either incorporating or removing some ABS cells in appropriate regions. This was repeated several times until the frame 141 and its optimal shape appeared.

\section{Conclusions}
%
This article is devoted to the initial component of the integrated design of geometrically complex structures, in particular, to the simulation of a mechanical response in the conceptual phase of architectural design. It aims at maximum possible automation of structural behaviour assessment in the early stages of the design and results in economic and reliable exploration of designer's creativity. A simple, though effective methodology based on an open source interface that allows for interconnecting existing computer aided design and structural analysis engineering tools was introduced. Based on three illustrative case studies, it can be conjectured that:

\begin{enumerate}
  \setlength{\itemsep}{1pt}
  \setlength{\parskip}{-1pt}
  \setlength{\parsep}{-1pt}
  \item [$\bullet$] if the architectural model is created with respect to a subsequent structural analysis, the proposed process is robust and free of further user intervention;
  \item [$\bullet$] in the case of a complex model, certain simplifications to the model are required,
                    however, the structural analysis can be still performed without the need for structural engineer's interventions;
  \item [$\bullet$] on the contrary, collaboration with experts in structural analysis, numerical methods and programming is necessary when solving extraordinary and/or very large structures;
  \item [$\bullet$] significant time savings in communication between structural engineers and architects were achieved when solving all three benchmarks, no matter the complexity. For example, 20 modifications of the Geometric Death Frequency-141 model was made within 14 days.
\end{enumerate}

Finally, let us emphasise that our aim is not to replace a detailed structural assessment up to the extent required in the advanced stages of the project (building certificate and/or operating documentation) but to provide architects, designers and artists with a simple tool assisting in better understanding of structural behaviour.

\section*{Acknowledgements}
%
The authors thank Federico D\'{i}az for his involvement in software testing and providing us with GDF input data. We also gratefully acknowledge the endowment of The ministry of industry and trade of the Czech Republic under project FR-TI1/568 and the European Social Fund under Grant No.CZ.1.07/2.3.00/30.0005 of Brno University of Technology (Support for the creation of excellent interdisciplinary research teams at Brno University of Technology). Finally, we would like to thank Ji\v{r}\'{\i} \v{S}ejnoha from CTU in Prague for a careful review of the manuscript.

\end{document}